# A Cross-Layer Solution in Scientific Workflow System for Tackling Data Movement Challenge

Dong Dai[1], Robert Ross[2], Dounia Khaldi[3], Yonghong Yan[4], Matthieu Dorier[2], Neda Tavakoli[1], and Yong Chen[1]

[1] Computer Science Department, Texas Tech University, USA
[2] Mathematics and Computer Science Division, Argonne National Laboratory, USA
[3] Institute for Advanced Computational Science, Stony Brook University, USA
[4] Computer Science Department, Oakland University, USA

## I. EXTENDED ABSTRACT

Scientific applications in HPC environment are more com-plex and more data-intensive nowadays. Scientists usually rely on workflow system to manage the complexity: simply define multiple processing steps into a single script and let the work-flow systems compile it and schedule all tasks accordingly. Numerous workflow systems have been proposed and widely used, like Galaxy, Pegasus, Taverna, Kepler, Swift, AWE, etc., to name a few examples.

Traditionally, scientific workflow systems work with parallel file systems, like Lustre, PVFS, Ceph, or other forms of remote shared storage systems. As such, the data (including the intermediate data generated during workflow execution) need to be transferred back and forth between compute nodes and storage systems, which introduces a significant performance bottleneck on I/O operations. Along with the enlarging perfor-mance gap between CPU and storage devices, this bottleneck is expected to be worse.

Recently, we have introduced a new concept of Compute-on-Data-Path to allow tasks and data binding to be more efficient to reduce the data movement cost. To workflow systems, the key is to exploit the data locality in HPC storage hierarchy: if the datasets are stored in compute nodes, near the workflow tasks, then the task can directly access them with better performance with less network usage. Several recent studies have been done regarding building such a shared storage system, utilizing compute node resources, to serve HPC workflows with locality, such as Hercules [1] and WOSS [2] etc. In this research, we further argue that providing a compute-node side storage system is not sufficient to fully exploit data locality. A cross-layer solution combining storage system, compiler, and runtime is necessary. We take Swift/T [3], a workflow system for data-intensive applications, as a prototype platform to demonstrate such a cross-layer solution.

### A. Swift/T with Hercules

Swift/T is the latest release that implements Swift program-ming language in HPC context. It permits users to easily express the logic of many-task applications using the high-level Swift language. Hercules is a distributed in-memory store based on Memcached. Hercules offers Swift/T workers a shared storage with POSIX interface with data locality. Fig. 1 shows an overall architecture of Swift/T with Hercules. There are multiple components in such a system. First, there is a Swift compiler that turns Swift scripts submitted from users into intermediate (Turbine) code. Second, those codes are sent to multiple script engines to generate the real tasks. Third, a load balancer takes charge of scheduling those tasks into workers. The scheduler in most cases works in a first-come-first-serve way. At last, the workers run the task and can access data either from Hercules or remote parallel file systems.

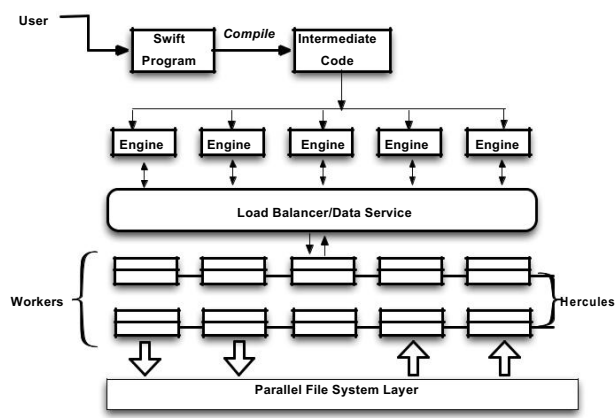

**Fig. 1:** Swift/T with Hercules architecture.

In Fig 2, we further show an example of Swift/T script and the directed acyclic task graph generated from such a script. The DAG is obtained from Swift/T compiler. With Hercules support, all the inputs, intermediate results, and outputs can be stored in compute nodes with better locality. Swift/T supports @location annotation to specify the location for a certain task, which allows to exploit the locality once programmers know exactly the location of data. However, it does not work well in the current system several reasons.

1) First, the compute-node side file system (Hercules) can not explicitly control data location nor expose the locality to users or applications. Scientists have to speculate the location and tell workflow engine to leverage the locality. This is not effective for real-world use cases.
2) Second, Swift compiler does not collect important meta-

data about data and tasks, like the size of intermediate datasets or the estimated execution time of a task. Without such information, it is hard for workflow scheduler to make better decisions.

3) Finally, after scheduling, Swift runtime does not provide a feedback of the scheduling to storage systems. The feedback actually can help the file system to re-organize or migrate their data to improve performance.

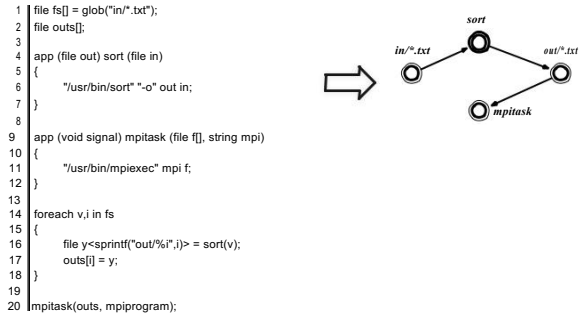

**Fig. 2:** Swift/T script and its task DAG.

It is clear to see, to tackle those challenges, only optimizing storage system or workflow engine is not sufficient. In this position poster, we argue that a cross-layer solution is needed.

### B. A Cross-Layer Solution

In this research, the proposed solution mainly contains three parts across multiple layers, i.e., file system, static compiler, and runtime scheduler.

**Location-Aware File System Extension.** We propose to extend the file system (Hercules in the current proof-of-concept evaluation platform) that is built on compute nodes with location-aware APIs. For instance, we extend OPEN API by adding a new mode argument (S_LOC) for the O_CREAT flag to denote the location we want this new file to be created at. If no such argument is provided, the file might be placed according to Hercules' current algorithm. In addition, the location metadata of a file is also stored in its extended attributes (xattr). Users or applications can use POSIX APIs like getxattr to retrieve it.

We also build a distributed location service for querying files and their locations. We allow users, applications, or runtime to explicitly specify arbitrary location for a file, which is real-loc. The real location of a file can be queried by clients through a distributed location metadata service.

**Hint-Assist Workflow Compiler.** We propose to extend the Swift/T language model with extra @ annotations to provide hints for workflow engine. New annotations include:

@size hints the size of an existing file.

@task hints the key parameters for a task. This should include the process number of it.
@compute-complex hints the estimated computation cost of a task. This is defined as a function of input data size. For example, @compute-complexity=@input means its computation cost is linear to the inputs.

@input-output-ratio hints the output size of a task based on its inputs. This is critical for compiler to estimate the size of generated intermediate results.

These hints are captured by Swift/T compiler. They will be attached to the generated task DAG. Such 'rich' metadata about tasks and datasets are critical for static analysis in compiler. It allows us to sort the DAG topologically to determine the earliest start time of each task and help runtime scheduler.

**Locality-Aware Workflow Scheduler.** In general, DAG-based workflow scheduling is an NP hard problem. In this research, we propose a heuristic strategy. We calculate a score for each ready task as its priority used in Swift/T load balancer. It covers the priority from compiler and also data movement cost. Specifically, it first calculates the length of the longest path from the final task to current task. Longer distance usually indicates a higher priority as it may slow down the overall workflow performance. Second, it further considers the dynamic available workers and the data movement cost for task $t_i$ to run on node $n_j$.

In addition to this heuristic scheduling, we further introduce another algorithm, called proactive scheduling to collaborate with file systems. Specifically, it will pre-schedule the non-ready tasks in DAG before their inputs are ready. The task might be pre-scheduled even only parts of its inputs are ready. The scheduling is also based on the previous heuristic score except the data movement cost is estimated and not accurate. The key advantage of such scheduling is that, based on this pre-scheduling decision, we can tell the file system to start pipelining the data to the target server. Hence, when the task is ready to run, its data will already be there, significantly reducing I/O time.

### C. Summary

In this research, we identify the challenges of exploit data locality in scientific workflow systems. We further propose a cross-layer solution, combining file system, compiler, and runtime together, to tackle these challenges.

### D. Acknowledge

This material is based upon work supported by the National Science Foundation under grant CCF-1409946 and CNS-1338078.

### REFERENCES

[1] F. R. Duro, J. G. Blas, F. Isaila, J. Carretero, J. Wozniak, and R. Ross, "Exploiting Data Locality in Swift/T Workflows Using Hercules," in Proc. NESUS Workshop, 2014.

[2] S. Al-Kiswany, E. Vairavanathan, L. B. Costa, H. Yang, and M. Ripeanu, "The Case for Cross-layer Optimizations in Storage: A Workflow-optimized Storage System," arXiv preprint arXiv:1301.6195, 2013.

[3] J. M. Wozniak, M. Wilde, and I. T. Foster, "Language Features for Scalable Distributed-Memory Dataflow Com-puting," in Data-Flow Execution Models for Extreme Scale Computing (DFM), 2014 Fourth Workshop on. IEEE, 2014, pp. 50–53.